\documentclass{article}

\newcommand{\BEAS}{\begin{eqnarray*}}
\newcommand{\EEAS}{\end{eqnarray*}}
\newcommand{\BEA}{\begin{eqnarray}}
\newcommand{\EEA}{\end{eqnarray}}
\newcommand{\BEQ}{\begin{equation}}
\newcommand{\EEQ}{\end{equation}}
\newcommand{\BIT}{\begin{itemize}}
\newcommand{\EIT}{\end{itemize}}
\newcommand{\BNUM}{\begin{enumerate}}
\newcommand{\ENUM}{\end{enumerate}}

\newcommand{\BA}{\begin{array}}
\newcommand{\EA}{\end{array}}

\newcommand{\reals}{{\mbox{\bf R}}}

\newcommand{\symm}{{\mbox{\bf S}}}  

\newcommand{\Tr}{\mathop{\bf Tr}}

\def\id{\mathbf{\mathbf{I}}}

\newcommand{\BP}{\begin{pmatrix}}
\newcommand{\EP}{\end{pmatrix}}
\newcommand{\mysubsection}[1]{\paragraph{#1.}}

\usepackage{fullpage,times,amsmath}
\usepackage{amsfonts}
\usepackage{amssymb}
\usepackage{psfrag}
\usepackage{graphics}
\usepackage{epsfig}

\begin{document}

\title{Sparse Covariance Selection via\\ Robust Maximum Likelihood Estimation}

\author{Onureena Banerjee\thanks{EECS Department, UC Berkeley, Berkeley, CA
94720. \texttt{onureena@eecs.berkeley.edu}}, Alexandre
d'Aspremont\thanks{ORFE Department, Princeton University,
Princeton, NJ 08544. \texttt{aspremon@princeton.edu}}, Laurent El
Ghaoui\thanks{EECS Department, UC Berkeley, Berkeley, CA 94720.
\texttt{elghaoui@eecs.berkeley.edu}}}

\maketitle

\begin{abstract}
We address a problem of covariance selection, where we seek a trade-off between a high likelihood
against the number of non-zero elements in the inverse covariance matrix. We solve a maximum
likelihood problem with a penalty term given by the sum of absolute values of the elements of the
inverse covariance matrix, and allow for imposing bounds on the condition number of the solution.
The problem is directly amenable to now standard interior-point algorithms for convex
optimization, but remains challenging due to its size. We first give some results on the
theoretical computational complexity of the problem, by showing that a recent methodology for
non-smooth convex optimization due to Nesterov can be applied to this problem, to greatly improve
on the complexity estimate given by interior-point algorithms. We then examine two practical
algorithms aimed at solving large-scale, noisy (hence dense) instances: one is based on a
block-coordinate descent approach, where columns and rows are updated sequentially, another
applies a dual version of Nesterov's method.
\end{abstract}

\section{Introduction}
Consider a data set with $n$ variables, drawn from a multivariate Gaussian distribution ${\cal
N}(0, \Sigma)$, where the covariance matrix $\Sigma$ is unknown.  When the number of variables $n$
is large, estimating the entries of $\Sigma$ becomes a significant problem.

In 1972, Dempster \cite{dempster1972} suggested reducing the number of parameters to be estimated
by setting to zero some elements of the inverse covariance matrix $\Sigma^{-1}$. This idea, known
as covariance selection, can lead to a more robust estimate of $\Sigma$ if enough entries in its
inverse are set to zero. Furthermore, conditional independence properties of the distribution are
determined by the locations of zeros in $\Sigma^{-1}$.  Hence the approach can be used to
simultaneously determine a robust estimate of the covariance matrix and, perhaps more importantly,
discover structure, namely conditional independence properties, in the underlying graphical model,
which is a useful information in its own right. Specific applications of covariance selection
include speech recognition \cite{bilmes1999, chen1999, bilmes2000} and gene expression data
analysis \cite{Dobr04,Dobr04a}.

In \cite{bilmes2000}, Bilmes proposed a method for covariance selection based on choosing
statistical dependencies according to conditional mutual information computed using training data.
Other recent work involves identifying those Gaussian graphical models that are best supported by
the data and any available prior information on the covariance matrix.  This approach is used by
\cite{jones2004, Dobr04} and is applied to gene expression data. Recently, \cite{Huan05,Dahl05}
considered penalized maximum likelihood estimation and \cite{Dahl05} in particular propose a set
of large scale methods to solve \emph{sparse} problems.

In this paper we focus on the problem of computing a sparse estimate of the covariance matrix
using only a large-scale, a priori \emph{dense} and noisy sample covariance matrix $\Sigma$.  Our
approach is based on $l_1$-penalized maximum likelihood, and can be interpreted as a "robust
maximum likelihood" method, where we assume that the true covariance matrix is within a
component-wise bounded perturbation of the sample one, and the estimate is chosen to maximize the
worst-case (minimal) likelihood. One of our goals is to provide an efficient algorithm to discover
structure, rather than solve problems where the inverse covariance matrix has an already known
sparse structure, as in \cite{Dahl05}.

Our contributions are as follows: we specify the problem and outline some of its basic properties
(section \ref{s:preliminaries}); we describe how one can apply a recent methodology for convex
optimization due to Nesterov \cite{nesterov2003}, and obtain as a result a computational
complexity estimate that has a much better dependence on problem size than interior-point
algorithms (section \ref{s:complexity}); we present two algorithms for solving large dense
problems (section \ref{s:algos}): a version of Nesterov's method applied to the dual problem, and
a block coordinate descent method. In section \ref{s:numericalresults} we present the results of
some numerical experiments comparing these two algorithms.

\section{Preliminaries}
\label{s:preliminaries}

\mysubsection{Problem setup}  For a given covariance matrix $\Sigma\in\symm_+^n$, and reals
numbers $0 \le \alpha < \beta$, our problem is formulated as
\begin{equation}\label{eq:sparseml-primal}
    p^\ast := \max_X \{ \log \det X - \langle \Sigma, X \rangle -
    \rho\|X\|_1 : \alpha \id_n \preceq X \preceq \beta \id_n\}
\end{equation}
with variable $X \in \symm^n$, and $\|X\|_1 :=  \sum_{i,j=1}^n |X_{ij}|$.  The parameter $\rho> 0$
controls the size of the penalty, hence the sparsity of the solution. Here $\langle \Sigma,X
\rangle = \Tr \Sigma X$ denotes the scalar product between the two symmetric matrices $\Sigma,X$.
The penalty term involving the sum of absolute values of the entries of $X$ is a proxy for the
number of its non-zero elements, and is often used in regression techniques, such as LASSO in
\cite{Tibs96}, when sparsity of the solution is a concern. In our model, the bounds
$(\alpha,\beta)$ on the eigenvalues of $X$ are fixed, and user-chosen; although we allow $\alpha =
0$, $\beta = +\infty$ in our model, such bounds are useful in practice to control the condition
number of the solution.

For $\rho = 0$, and provided $\Sigma \succ 0$, problem (\ref{eq:sparseml-primal}) has a unique
solution $X^\ast = \Sigma^{-1}$, and the corresponding maximum-likelihood estimate $\hat{\Sigma} =
\Sigma$. Due to noise in the data, in practice, the sample estimate $\Sigma$ may not have a sparse
inverse, even if the underlying graphical model exhibits conditional independence properties; by
striking a trade-off between maximality of the likelihood and number of non-zero elements in the
inverse covariance matrix, our approach is potentially useful at \emph{discovering structure},
precisely conditional independence properties in the data.  At the same time, it serves as a
regularization technique: when $\Sigma$ is rank-deficient, there is no well-defined
maximum-likelihood estimate, whereas it can be shown that the solution to problem
(\ref{eq:sparseml-primal}) is always unique and well-defined for $\rho>0$, even if $\alpha = 0$
and/or $\beta = \infty$.

\mysubsection{Robustness, duality, and bounds} \label{ss:robustness} In the case when $\alpha =0$,
$\beta = \infty$, we write (\ref{eq:sparseml-primal}) as
\[
\max_{X \succ 0} \: \min_{\Vert U \Vert_{\infty} \leq \rho} \: \log \det X - \langle X, \Sigma+U
\rangle ,
\]
where $\|U\|_\infty$ denotes the maximal absolute value of the entries of $U$.  This corresponds
to seeking an estimate with maximal worst-case likelihood, over all componentwise bounded additive
perturbations $\Sigma+U$ of the sample covariance matrix $\Sigma$. Such a "robust optimization"
interpretation can be given to a number of estimation problems, most notably support vector
machines for classification.

The above leads to the following equivalent, dual problem (with $\alpha =0$, $\beta = \infty$):
\begin{equation}\label{eq:sparseml-dual}
    d^\ast := \min \: -\log \det (\Sigma+U)- n : \| U \|_{\infty} \leq \rho, \;\; \Sigma+U\succ 0 .
\end{equation}
Note that the diagonal elements of an optimal $U$ are simply $U_{ii} = \rho$. The corresponding
covariance matrix estimate is $\hat{\Sigma} := \Sigma+U$.  Since the above dual problem has a
compact feasible set, both primal and dual problems are equivalent.

In the case when $\alpha = 0$, $\beta = \infty$, we can derive finite, a priori bounds on the
condition number of the solution.  Indeed, we can show that we can always assume $\alpha(n) \id_ n
\preceq X \preceq \beta(n) \id_n$, where $\alpha(n) := 1/(\|\Sigma\|+n\rho)$ and $\beta(n) =
n/\rho$. These bounds guarantee strict convexity of the objective, as well as existence,
boundedness and uniqueness of the solution when $\rho>0$, even if no bounds are set a priori.

\section{Complexity}
\label{s:complexity}

\paragraph{First- vs.\ second-order methods.}
Of course, problem (\ref{eq:sparseml-primal}) is convex and can readily be solved using
interior point methods (see \cite{Boyd03} for example). However, such second-order methods become
quickly impractical for solving (\ref{eq:sparseml-primal}), since the corresponding complexity to
compute an $\epsilon$-suboptimal solution is $O(n^6 \log (1/\epsilon))$. The authors of
\cite{Dahl05} developed interior-point algorithms to solve a problem related to
(\ref{eq:sparseml-primal}), where a (sparse) structure of the solution is known a priori. Here our
focus is on relatively large, dense problems, for which a solution of moderate accuracy is enough.
Note that we cannot expect to do better than $O(n^3)$, which is the cost of solving the
non-penalized problem for dense covariance matrices $\Sigma$. The recently-developed first-order
algorithms due to \cite{nesterov2003} trade-off a better dependence on problem size against a
worse dependence on accuracy, usually $1/\epsilon$ instead of its logarithm.  The method we
describe next has a complexity of $O(n^{4.5}/\epsilon)$. This is a substantial improvement over
interior-point methods when $\epsilon$ is not too small, which is often the case in practice.  In
addition, the memory space requirement of the first-order method is much lower than that of
interior-point methods, which involve storing a dense Hessian, and hence become quickly
prohibitive with a problem having $O(n^2)$ variables.

\paragraph{Nesterov's format.}
We can write (\ref{eq:sparseml-dual}) in the format given in \cite{nesterov2003}:
\[
\min_{X \in {\cal Q}_1} \: \max_{U \in {\cal Q}_2} \: \hat{f}(X) + \langle A(X),U \rangle =
\min_{X \in {\cal Q}_1} \: f(X) ,
\]
where we define $\hat{f}(X) = -\log \det X +  \langle \Sigma,X \rangle$, $A = \rho I_{n^2}$, and
${\cal Q}_1 := \left\{ X \in {\cal S}^n ~:~ \alpha \id_n \preceq X \preceq \beta \id_n \right\}$,
${\cal Q}_2 := \left\{ U \in {\cal S}^n  ~:~ \|U\|_\infty \leq 1 \right\}$.

\paragraph{Prox-functions and related parameters.}
To ${\cal Q}_1$ and ${\cal Q}_2$ we now associate norms and so-called prox-functions. For ${\cal
Q}_1$, we use the Frobenius norm, and a prox-function
\[
d_1(X) = -\log \det X + \log \beta .
\]
The function $d_1$ is strongly convex on ${\cal Q}_1$, with a convexity parameter of $\sigma_1 =
1/\beta^2$, in the sense that $\nabla^2 d_1(X)[H,H] = \Tr (X^{-1}H X^{-1}H) \ge \beta^{-2}
\|H\|_F^2$ for every $H$. Furthermore, the center of the set, $X_0:=\arg\min_{X \in {\cal Q}_1} \:
d_1(X)$ is $X_0 = \beta \id_n$, and satisfies $d_1(X_0) = 0$. With our choice, we have $D_1 :=
\max_{X \in {\cal Q}_1} \: d_1(X) = n\log (\beta/\alpha)$.

To ${\cal Q}_2$, we also associate the Frobenius norm, and the prox-function $d_2(U) =
\|U\|_F^2/2$. With this choice, the center $U_0$ of ${\cal Q}_2$ is $U_0 = 0$. Furthermore, the
function $d_2$ is strictly convex on its domain, with convexity parameter with respect to the
$1$-norm $\sigma_1 = 1$, and we have $D_2 := \max_{U \in {\cal Q}_2} \: d_2(U) = n^2/2$.

The function $\hat{f}$ has a gradient that is Lipschitz-continuous with respect to the Frobenius
norm on the set ${\cal Q}_1$, with Lipschitz constant $M = 1/\alpha^2$. Finally, the norm (induced
by the Frobenius norm) of the operator $A$ is $\|A\| = \rho$.

\paragraph{Idea of the method.}
The method is based on replacing the objective of the original problem, $f(X)$, with
$f_\epsilon(X)$, where $\epsilon>0$ is the given desired accuracy, and $f_\epsilon$ is the
penalized function involving the prox-function $d_2$:
\begin{equation}\label{eq:feps-def}
    f_\epsilon(X) := \hat{f}(X) + \max_{U \in {\cal Q}_2} \langle X,U \rangle - (\epsilon/2D_2)
    d_2(U).
\end{equation}
The above function turns out to be a smooth uniform approximation to $f$ everywhere, with maximal
error $\epsilon/2$.  Furthermore, the function $f_\epsilon$ is Lipschitz-continuous, with
Lipschitz constant given by $L(\epsilon) := M+D_2\|A\|^2/(2 \sigma_2\epsilon)$. A specific
gradient algorithm for smooth, constrained convex minimization is then applied to the smooth
convex function $f_\epsilon$, with convergence rate in $O(\sqrt{L(\epsilon)/\epsilon})$.
Specifically, the algorithm is guaranteed to produce an $\epsilon$-suboptimal solution after a
number of steps not exceeding

\begin{equation}\label{eq:N-def}
N(\epsilon) := 4 \|A\| \sqrt{\displaystyle\frac{D_1D_2}{\sigma_1\sigma_2} } \cdot
\frac{1}{\epsilon} + \sqrt{\frac{MD_1}{\sigma_1\epsilon}} = \frac{\kappa \sqrt{n(\log
\kappa)}}{\epsilon} ( 4 n \alpha \rho + \sqrt{\epsilon})  .
\end{equation}

\paragraph{Nesterov's algorithm.} Choose $\epsilon>0$ and set $X_0 =  \beta\id_n$,
and proceed as follows.

\noindent {\bf For $k = 0,\ldots,N(\epsilon)$ do}
\begin{enumerate}
    \item Compute $\nabla f_\epsilon(X_k) = -X^{-1} + \Sigma + U^\ast(X_k)$, where $U^\ast(X)$
    solves (\ref{eq:feps-def}).
    \item Find $Y_k = \arg\min_{Y} \: \{\langle \nabla f_\epsilon(X_k) , Y-X_k \rangle +
\frac{1}{2} L(\epsilon) \|Y-X_k\|_F^2 ~:~ Y  \in {\cal Q}_1 \}$.
    \item Find $Z_k = \arg\min_X \left\{ \frac{L(\epsilon)}{\sigma_1} d_1(X) + \sum_{i=0}^k \frac{i+1}{2}
    \langle \nabla f_\epsilon(X_i),X-X_i \rangle ~:~ X \in {\cal Q}_1 \right\}$.
    \item Update $X_k = \frac{2}{k+3} Z_k + \frac{k+1}{k+3} Y_k$.
\end{enumerate}

\paragraph{Complexity estimate.}
For step $1$, the gradient of $f_\epsilon$ is readily computed in closed form, via the computation
of the inverse of $X$. Step $2$  essentially amounts to projecting on ${\cal Q}_1$, and requires
an eigenvalue problem to be solved; likewise for step $3$.  In fact, each iteration costs
$O(n^3)$. The number of iterations necessary to achieve an objective with absolute accuracy less
than $\epsilon$ is given in (\ref{eq:N-def}) by $N(\epsilon) = O(n/\epsilon)$ (if $\rho >0$).
Thus, the overall complexity when $\rho>0$ is $O(n^{4.5}/\epsilon)$, as claimed.

\section{Algorithms}
\label{s:algos} The algorithms presented next address problem (\ref{eq:sparseml-primal}) in the
case when no bounds are given a priori, that is, $\alpha = 0$ and $\beta = \infty$.  While they do
not share all the nice theoretical computational complexity properties of the algorithm presented
earlier, they seem to work well in practice.  We are currently implementing the algorithm in section \ref{s:complexity} and will include it in our experiments in a future version of this paper.

\mysubsection{A dual version of Nesterov's method} In our experiments, we have used a preliminary
version of Nesterov's method with $\alpha = 0$ and $\beta = \infty$ that, instead of working on the primal (\ref{eq:sparseml-primal}),
addresses the dual (\ref{eq:sparseml-dual}). The algorithm is essentially the same as that
presented in section \ref{s:complexity}, with the following setup. Define ${\cal Q}_1 := \{ x ~:~
U \in {\cal S}^n  ~:~ \|U\|_\infty \leq 1 \}$, ${\cal Q}_2 = \{ U \in {\cal S}_+^n ~:~ \Tr U \le
n/\rho \}$, and
\[
f(x) =  \max_{U \in {\cal Q}_2} \langle A(X),U \rangle - \hat{\phi}(U) , \;\; A = -\id_{n^2}, \;\;
\hat{\phi}(u) = -\log\det U + \langle \Sigma,U \rangle .
\]
To simplify the projections required by the algorithm, we make the assumption that $\rho$ is small
enough to ensure that ${\cal Q}_1$ is entirely included in the cone of positive semidefinite
matrices.  In practice, we have found that the algorithm does well, even when this condition does
not hold. As in section \ref{s:complexity}, each step of this algorithm costs $O(n^3)$ operations.

To ${\cal Q}_1$, we associate the Frobenius norm in $\reals^{n \times n}$, and a prox-function
defined for $x \in {\cal Q}_1$ by $d_1(X) = (1/2) \|X\|_F^2$; the center of ${\cal Q}_1$ with
respect to $d_1$ is $X_0 = 0$, and $D_1 = \max_{X \in {\cal Q}_1} \: d_1(X) = \rho^2 n^2 /2$.
Furthermore, the function $d_1$ is strictly convex on its domain, with convexity parameter with
respect to the Frobenius norm $\sigma_1 = 1$. For ${\cal Q}_2$ we use the dual of the standard
matrix norm (denoted $\|\cdot\|^\ast_2$), and a prox-function $d_2(U) = \Tr (U \log U ) + \log n$,
where $\log$ refers to the {\em matrix} (and not componentwise) logarithm. The center of the set
${\cal Q}_2$ is $U_0 = n^{-1}\id_n$, and $d_2(U_0) = 0$. We have $\max_{u \in {\cal Q}_2} \:
d_2(u) \leq \log n := D_2$. The convexity parameter of $d_2$ on its domain with respect to
$\|\cdot\|_2^\ast$, is bounded below by $\sigma_2 = 1/2$ \cite{bental2004}. Finally, the norm of
the operator $A$, with respect to the Frobenius norm and the dual of the standard norm, is $1$.

\mysubsection{Dual block-coordinate descent} \label{ss:blkcoordesc} We now describe an algorithm
based on block-coordinate descent, which solves the dual problem (\ref{eq:sparseml-dual}) by
optimizing over one column/row pair at a time. For simplicity, we consider the case when
$\alpha=0$ and $\beta = \infty$, which corresponds to providing no a priori bounds on the
condition number of the solution.

The algorithm is initialized with the estimate $\hat{\Sigma} = \Sigma+\rho I \succ 0$, since the
optimal diagonal values of $U$ in (\ref{eq:sparseml-dual}) are simply $\rho$. To update
$\hat{\Sigma}$, we solve (\ref{eq:sparseml-dual}) where all elements in $U$, except the
off-diagonal values of one column (and corresponding row), are fixed. Using a permutation of rows
and columns, we can always express the problem as one of optimizing over the last column/row pair:
\begin{equation}\label{eq:bcd-gen}
    \min_{u,v} \: \log \det
\begin{pmatrix}
\hat{\Sigma_{11}} & \hat{\Sigma}_{12}+u \\
\ast & \hat{\sigma}_{22}\end{pmatrix} ~:~ |u| \le \rho,
\end{equation}
where inequalities are understood componentwise, and the current estimate $\hat{\Sigma}$ of the
covariance matrix is partitioned into four blocks, with $\hat{\Sigma}_{11}$ the upper-left
$(n-1)\times(n-1)$ block, and $(\hat{\Sigma}_{12},\hat{\sigma}_{22})$ the last column. The
sub-problem (\ref{eq:bcd-gen}) reduces to a simple box-constrained quadratic program.
Specifically, the corresponding column/row $\hat{\Sigma}_{12}$ of the current estimate
$\hat{\Sigma}$ is updated with $\hat{\Sigma}_{12}+u^\ast$, where
\begin{equation}\label{eq:box-QP}
    u^\ast = \arg\min_{u} \: u^T\hat{\Sigma}_{11}^{-1}u ~:~ |u| \le \rho .
\end{equation}

The stopping criterion we can use involves checking if the primal-dual gap of the problem
(\ref{eq:sparseml-dual}) is less than a given tolerance $\epsilon$, which translates as $\langle
\Sigma,X \rangle + \rho \|X\|_1 \le n+\epsilon$, where $X =\hat{C}^{-1}$.  In practice though, we
have found that a fixed number of sweeps (say, $K = 4$) through all column/row pairs is usually
enough.

It can be proven that the algorithm converges, however we do not know the complexity of the
algorithm.  A simple modification of it, where we impose a condition number constraint on the
estimate $\hat{\Sigma}$ at each column/row update, can be shown to converge in $O(n^6)$. Hence,
the dual BCD algorithm is not really competitive from the theoretical complexity point of view,
but the version given here remains attractive because it is simple to implement, and does well in
practice.  Since each column/row update costs $O(n^3)$, when only a few, fixed number $K$ of
sweeps through all columns is done, the cost is $O(Kn^4)$.

\section{Numerical Results}
\label{s:numericalresults} 
\begin{figure}[ht]
\begin{center}
\psfrag{ap}[c][c]{Noisy inverse $\Sigma^{-1}$} \psfrag{bp}[c][c]{Solution for $\rho=\sigma$}
\psfrag{cp}[c][c]{Original inverse $A$}
\includegraphics[width= \textwidth]{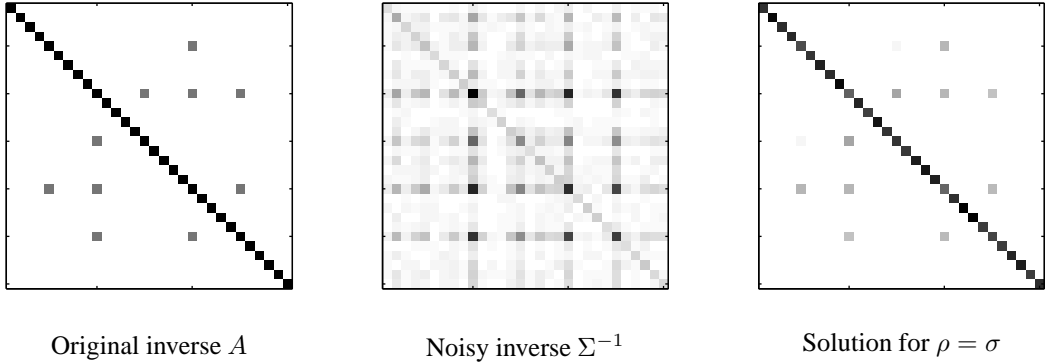}
\end{center}
\caption{\label{fig:pattern} Recovering the sparsity pattern. We plot the original inverse covariance
matrix $A$, the noisy inverse $\Sigma^{-1}$ and the solution to problem
(\ref{eq:sparseml-primal}) for $\rho=.13$.}
\end{figure}

\mysubsection{Recovering structure} To illustrate our method we
consider a synthetic example constructed by adding noise to a covariance matrix which has a sparse
inverse. Precisely, we take a random $n \times n$ sparse matrix $A$, with $n=30$, and then add a
uniform noise of magnitude $\sigma=.13$ to its inverse to form our matrix $\Sigma$. Our concern
here is to check if our methods reach a degree of precision sufficient to recover the problem
structure, despite the noise masking it. Figure \ref{fig:pattern} shows the original sparse matrix
$A$, the noisy inverse $\Sigma^{-1}$ and the solution $X$ of problem (\ref{eq:sparseml-primal})
solved using Nesterov's algorithm detailed in section \ref{s:algos} with parameter $\rho=\sigma$.

\begin{figure}[ht]
\begin{center}
\begin{tabular}{cc}
\psfrag{rho}[t][b]{$\rho/\sigma$}
\psfrag{coefs}[b][t]{\small{Coefficients}}
\psfrag{Aint}[c][c]{\small{V}}
\psfragscanon
\includegraphics[width=.48\textwidth]{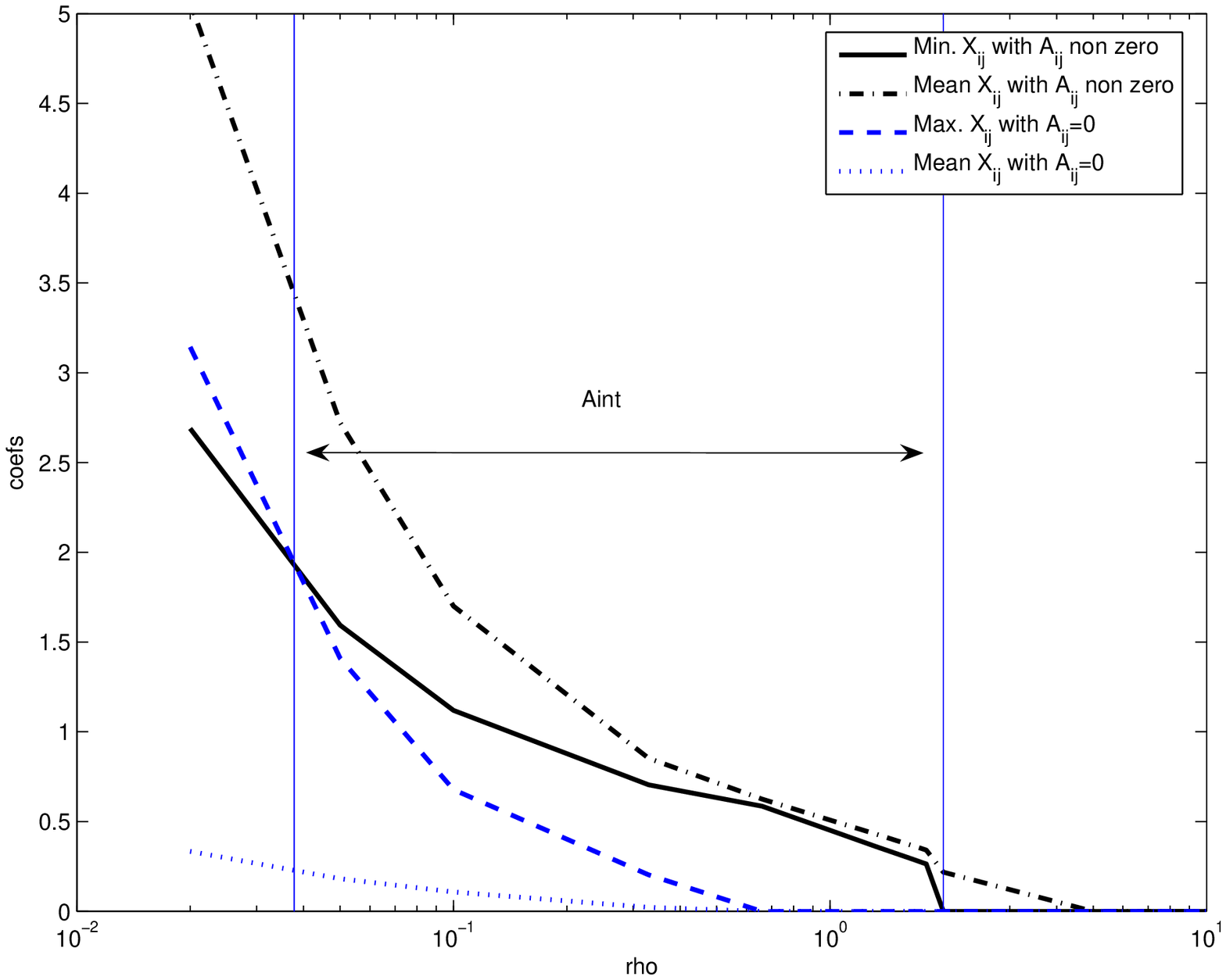}
\psfrag{rhob}[t][b]{$\log(\rho/\sigma)$}
\psfrag{err}[b][t]{\small{Error (in \%)}}
\includegraphics[width=.48 \textwidth]{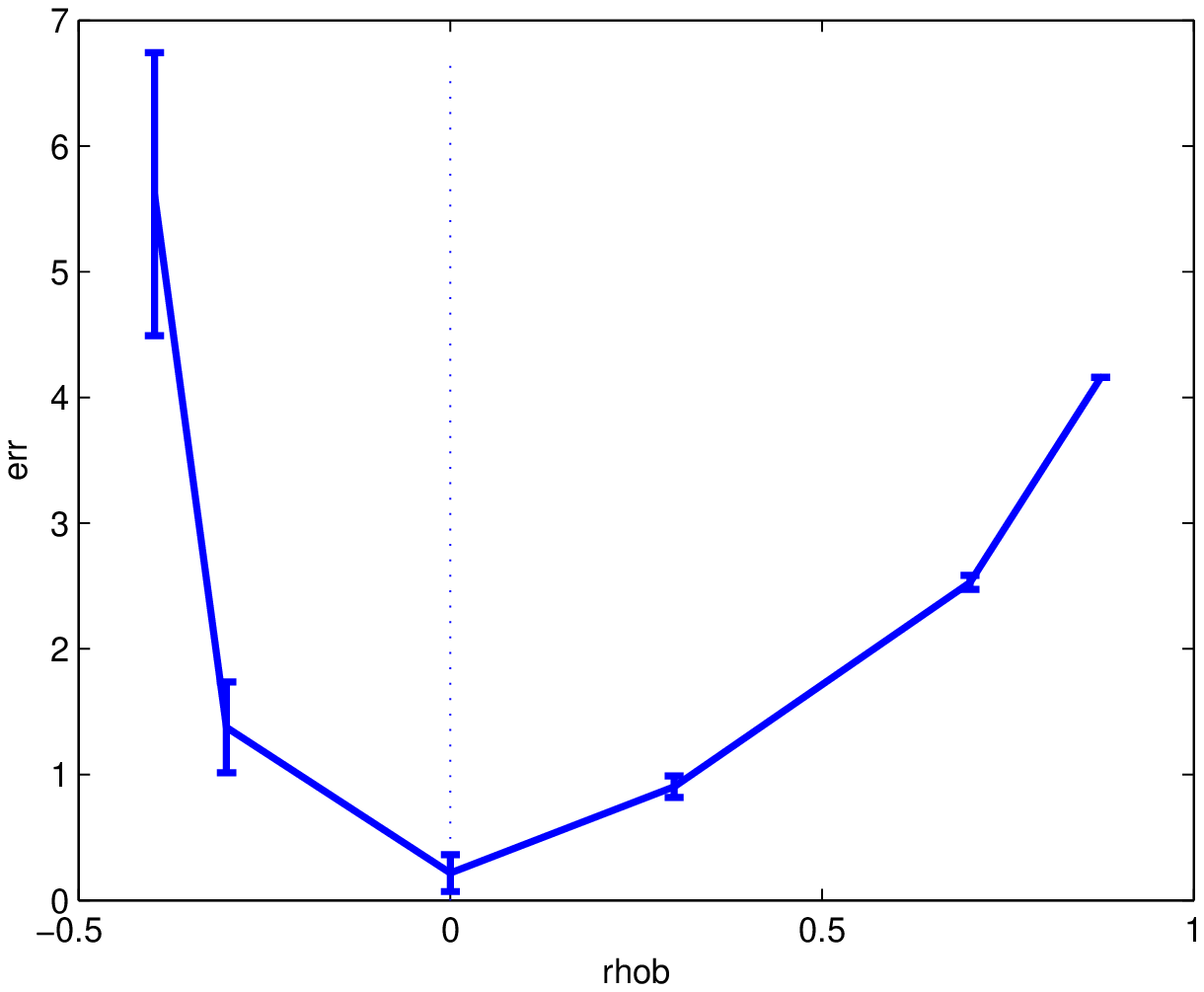}
\end{tabular}
\caption{\label{fig:rhoVSsigma} Recovering structure. \emph{Left}:
zero and nonzero matrix coefficients versus $\rho$. \emph{Right}:
average and standard deviation of the percentage of errors (false
positives + false negatives) versus $\rho$ on random problems.}
\end{center}
\end{figure}

Of course, the simple example above assumes that we set $\rho$ equal to the noise level $\sigma$,
so we look at what happens when the value of $\rho$ is set above or below $\sigma$. In Figure
\ref{fig:rhoVSsigma} on the \emph{left}, we solve problem (\ref{eq:sparseml-primal}) for various
values of $\rho$. We plot the minimum $\min\{X_{ij}:A_{ij}\ne 0\}$ (solid) and mean (dash-dot)
magnitude of those coefficients in the solution $X$ corresponding to non zero coefficients in $A$
(solid line) against the maximum $\max\{X_{ij}:A_{ij}=0\}$ (dashed line) and mean (dotted line)
magnitude of coefficients in $X$ corresponding to zeros in $A$ (we only consider off-diagonal
coefficients). For all values of $\rho$ within the interval $V$ shown in Figure
\ref{fig:rhoVSsigma}, the minimum is larger than the maximum, meaning that for all thresholding
levels between the minimum and maximum, we exactly recover the original matrix $A$. As we can
observe, the range $V$ of values of $\rho$ for which this happens is fairly large and so is the
gap between the minimum and maximum within the interval $V$.

In Figure \ref{fig:rhoVSsigma} on the \emph{right}, for various values of $\rho$, we randomly
sample 10 noisy matrices $\Sigma$ with $n=50$ and $\sigma=.1$ and compute the number of
misclassified zeros and nonzero elements in the solution to (\ref{eq:sparseml-primal}) produced by
the block coordinate descent method. We plot the average percentage of errors (number of elements
incorrectly set to zero plus number of elements incorrectly kept nonzero, divided by $n^2$), as
well as error bars corresponding to one standard deviation.

\begin{figure}[f]
\begin{center}
\begin{tabular}{cc}
\psfrag{cputime}[b][t]{\small{CPU Time (in seconds)}}
\psfrag{nsize}[t][b]{\small{Problem Size $n$}}
\includegraphics[width=0.48\textwidth]{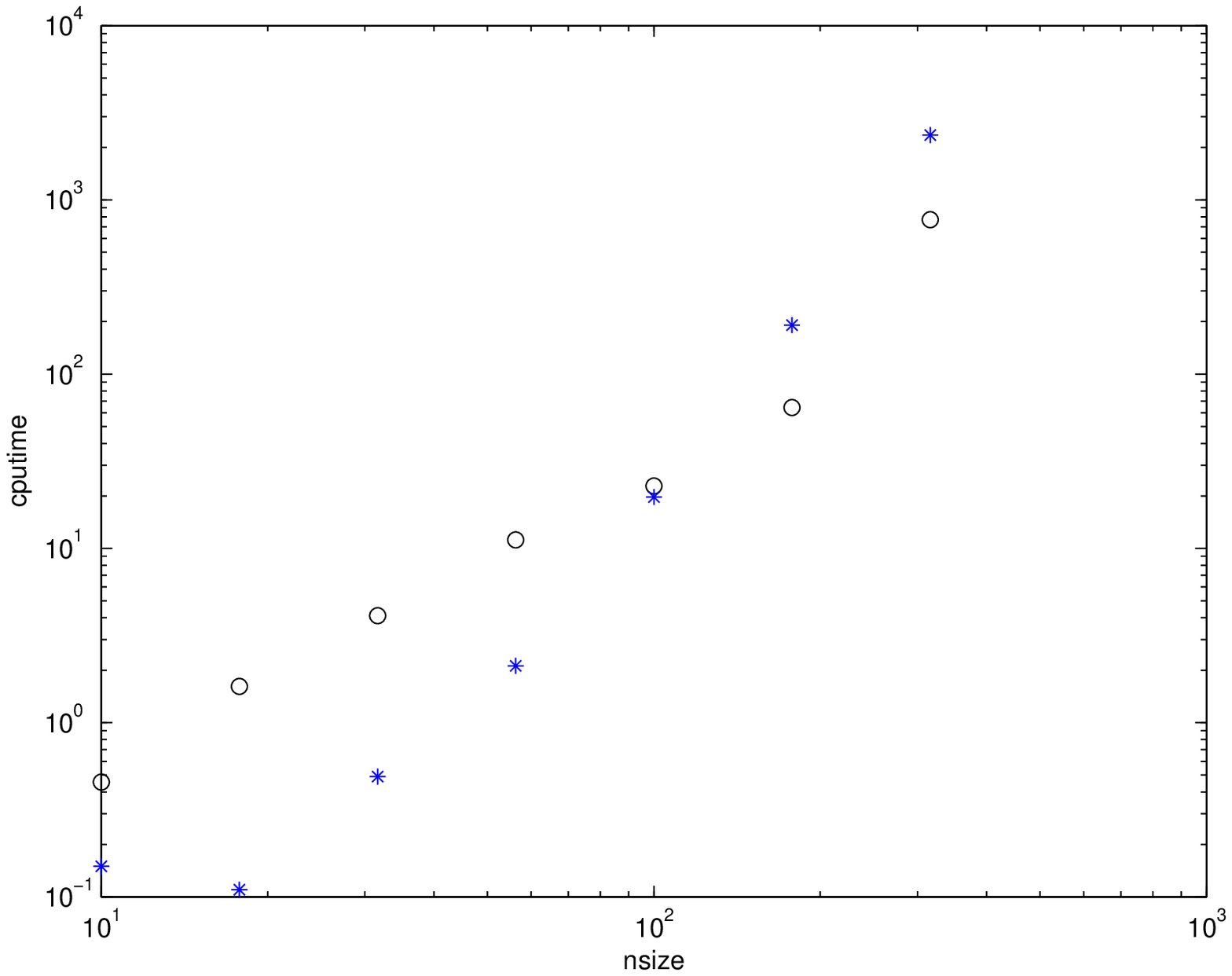}&
\psfrag{cputime}[t][b]{\small{CPU Time (in seconds)}}
\psfrag{optimality}[b][t]{\small{Duality Gap}}
\psfrag{A}[c][c]{\small{A}}
\psfrag{B}[c][c]{\small{C}}
\psfrag{C}[c][c]{\small{B}}
\includegraphics[width=0.48 \textwidth]{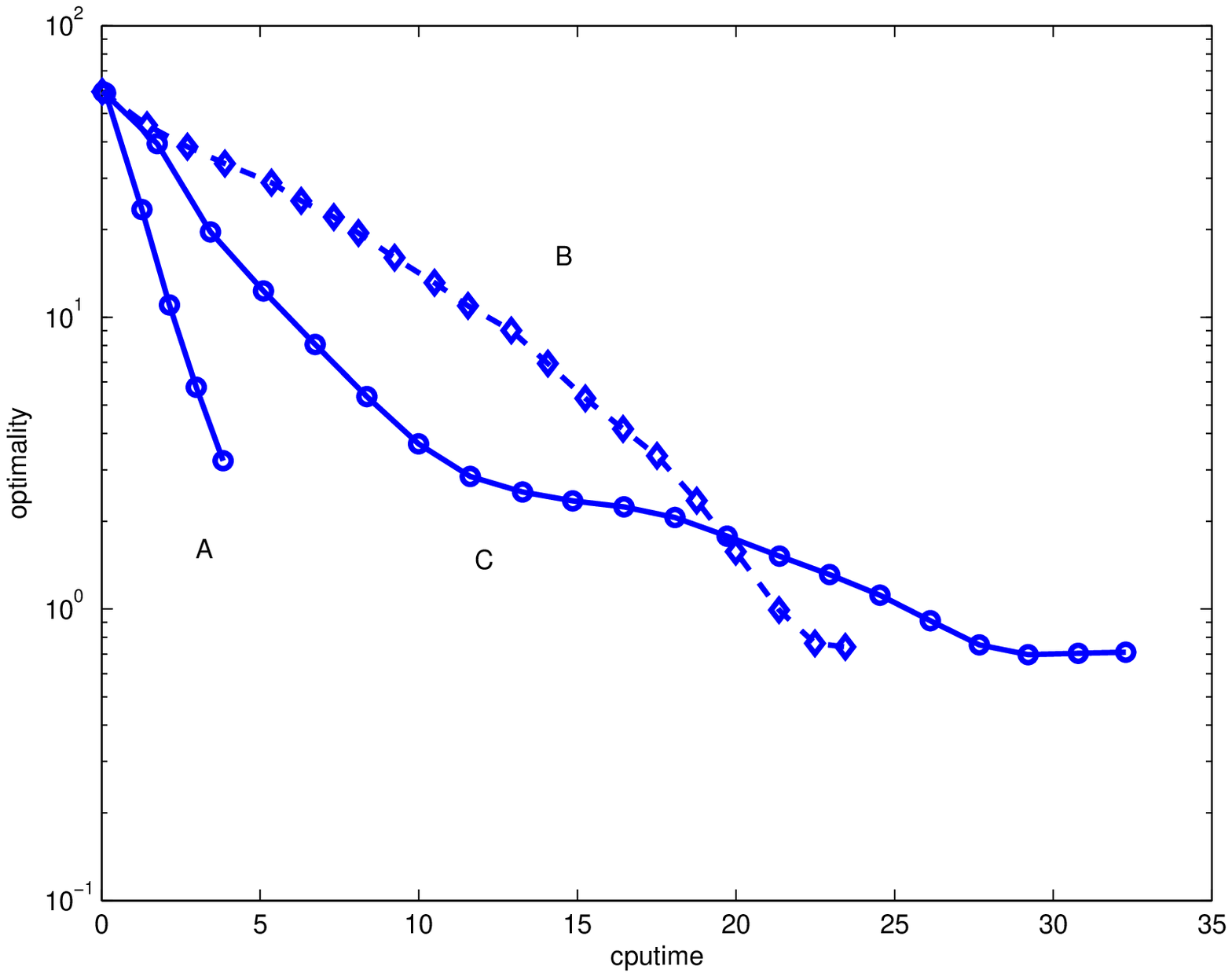}
\end{tabular}
\caption{\label{fig:cputime} Computing time. \emph{Left}: we plot CPU time (in seconds) to reach
gap of $\epsilon=1$ versus problem size $n$ on random problems, solved using Nesterov's method
(stars) and the coordinate descent algorithm (circles). \emph{Right}: convergence plot a random
problem with $n=100$, this time comparing Nesterov's method where $\epsilon=5$ (solid line A) and
$\epsilon=1$ (solid line B) with one sweep of the coordinate descent method (dashed line C).}
\end{center}
\end{figure}

\begin{figure}[f]
\begin{center}
\begin{tabular}{cc}
\psfrag{coefs}[t][b]{\small{Coefficients}}
\psfrag{magnitude}[b][t]{\small{Magnitude}}
\includegraphics[width=0.48\textwidth]{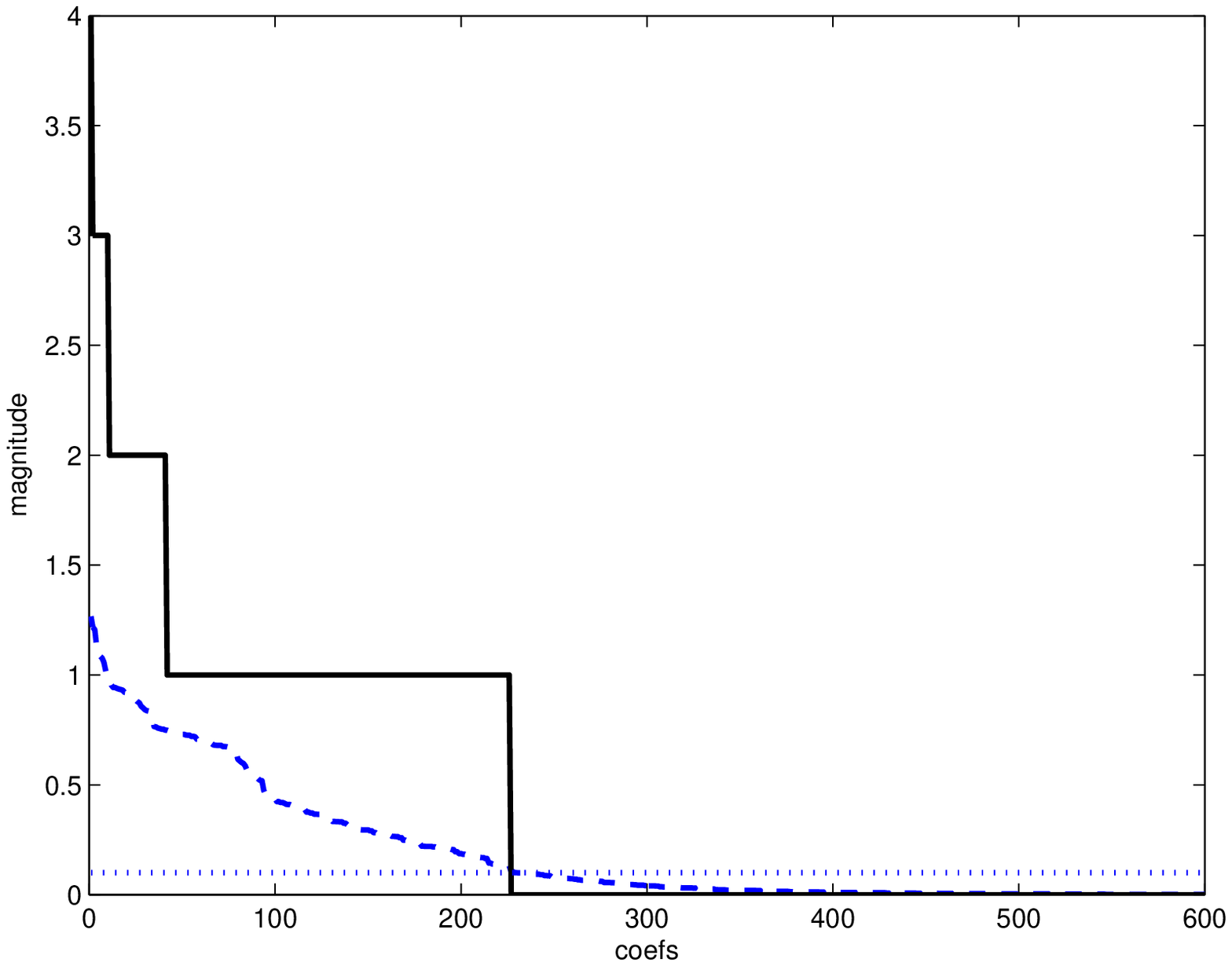}&
\psfrag{coefs}[t][b]{\small{Coefficients}}
\psfrag{magnitude}[b][t]{\small{Magnitude}}
\includegraphics[width=0.48 \textwidth]{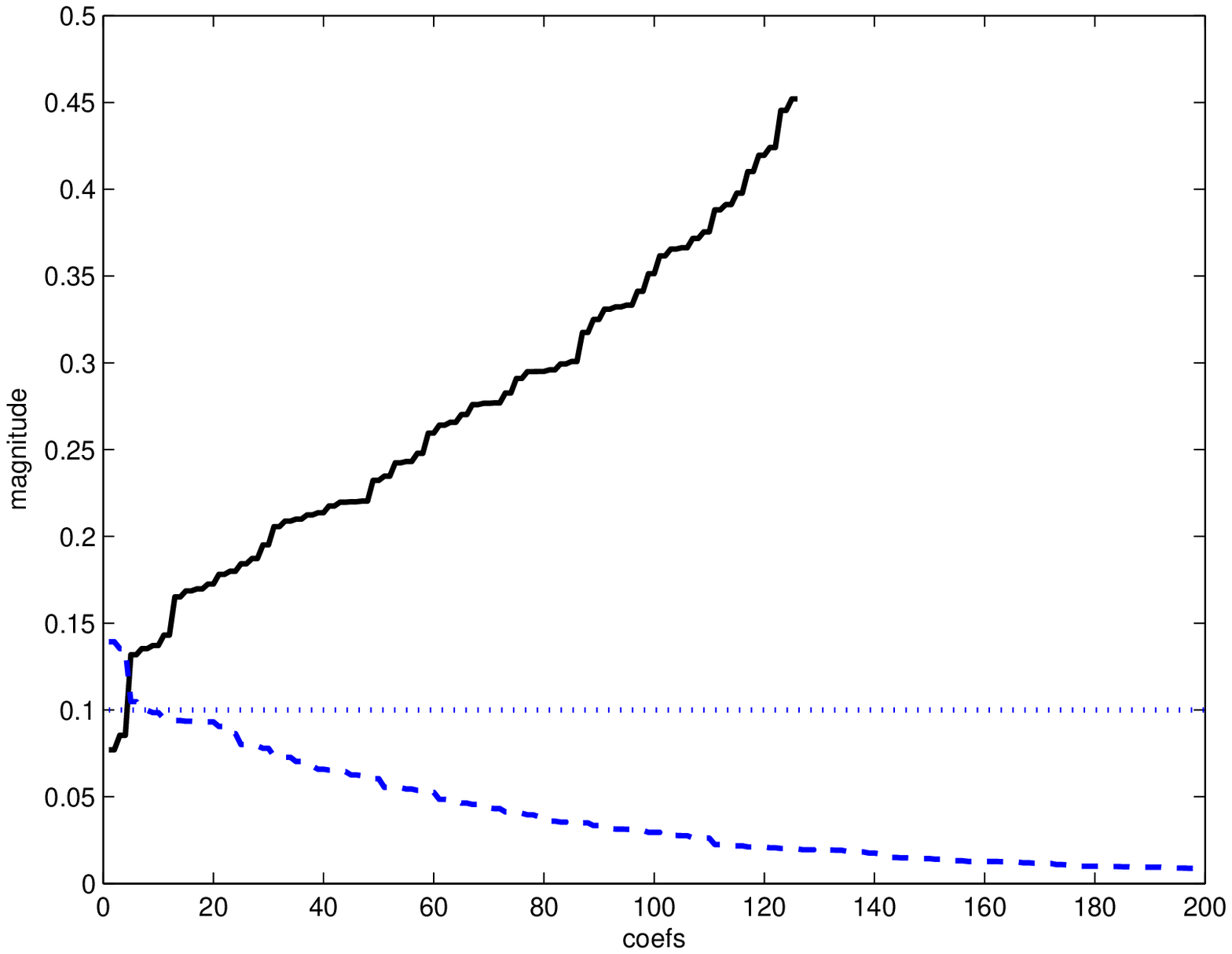}
\end{tabular}
\caption{\label{fig:error} Classification Error. \emph{Left}: we plot the coefficient magnitudes
for the original matrix $A$ (solid line) and the solution (dashed line) in decreasing order. The
dotted line is at the signal to noise level $\sigma$. \emph{Right}: we plot the magnitude of those
coefficients in the solution associated with nonzero elements in $A$ (solid line), ranked in
increasing magnitude, together with the magnitude of those coefficients in the solution associated
with zero elements in $A$ (dashed line), in decreasing order of magnitude. Again, the dotted line
is at the signal to noise level $\sigma$ and we only consider off-diagonal elements.}
\end{center}
\end{figure}

\mysubsection{Large-scale problems} We now compare Nesterov's and
coordinate descent algorithms on a set of randomly generated
examples. The noisy matrix $\Sigma$ is generated as above and we
plot in Figure \ref{fig:cputime} CPU time against problem size for
various $n$ and duality gap versus CPU time in seconds for a
random problem of size $100$ with a few nonzero coefficients. In
practice, a low-precision solution to (\ref{eq:sparseml-primal})
is sufficient to identify most of the nonzero coefficients in the
original matrix $A$. In Figure \ref{fig:error} we show the
classification error made by the solution on the example with
$n=100$. Typical computing time for a problem with $n=300$ is
about 20 minutes. (All CPU times computed on a 1.5Ghz PowerBook G4
laptop).

\newpage

\paragraph{Acknowledgments.} The authors would like to thank Francis Bach, Peter Bartlett and
Martin Wainwright for enlightening discussions on the topic.

\bibliographystyle{alpha}
\small{\bibliography{SparseMLArXiv}}

\newcommand{\etalchar}[1]{$^{#1}$}
\begin{thebibliography}{DHJ{\etalchar{+}}04}

\bibitem[Bil99]{bilmes1999}
J.~A. Bilmes.
\newblock Natural statistic models for automatic speech recognition.
\newblock {\em Ph.D. thesis, UC Berkeley, Dept. of EECS, CS Division}, 1999.

\bibitem[Bil00]{bilmes2000}
J.~A. Bilmes.
\newblock Factored sparse inverse covariance matrices.
\newblock {\em IEEE International Conference on Acoustics, Speech, and Signal
  Processing}, 2000.

\bibitem[BTN04]{bental2004}
A.~Ben-Tal and A.~Nemirovskii.
\newblock Non-euclidean restricted memory level method for large-scale convex
  optimization.
\newblock {\em MINERVA Working paper}, 2004.

\bibitem[BV04]{Boyd03}
S.~Boyd and L.~Vandenberghe.
\newblock {\em Convex Optimization}.
\newblock Cambridge University Press, 2004.

\bibitem[CG99]{chen1999}
S.~S. Chen and R.~A. Gopinath.
\newblock Model selection in acoustic modeling.
\newblock {\em EUROSPEECH}, 1999.

\bibitem[Dem72]{dempster1972}
A.~P. Dempster.
\newblock Covariance selection.
\newblock {\em Biometrics}, 28(1):157--75, 1972.

\bibitem[DHJ{\etalchar{+}}04]{Dobr04a}
A.~Dobra, C.~Hans, B.~Jones, J.R. J.~R. Nevins, G.~Yao, and M.~West.
\newblock Sparse graphical models for exploring gene expression data.
\newblock {\em Journal of Multivariate Analysis}, 90(1):196--212, 2004.

\bibitem[DRV05]{Dahl05}
J.~Dahl, V.~Roychowdhury, and L.~Vandenberghe.
\newblock Maximum likelihood estimation of gaussian graphical models: numerical
  implementation and topology selection.
\newblock {\em UCLA preprint}, 2005.

\bibitem[DW04]{Dobr04}
A.~Dobra and M.~West.
\newblock Bayesian covariance selection.
\newblock {\em Working paper, ISDS, Duke University}, 2004.

\bibitem[HLP05]{Huan05}
J.~Z. Huang, N.~Liu, and M.~Pourahmadi.
\newblock Covariance selection and estimattion via penalized normal likelihood.
\newblock {\em Wharton Preprint}, 2005.

\bibitem[JCD{\etalchar{+}}04]{jones2004}
B.~Jones, C.~Carvalho, C.~Dobra, A.~Hans, C.~Carter, and M.~West.
\newblock Experiments in stochastic computation for high-dimensional graphical
  models.
\newblock {\em ISDS Discussion Paper 04-01}, 2004.

\bibitem[Nes03]{nesterov2003}
Y.~Nesterov.
\newblock Smooth minimization of nonsmooth functions.
\newblock {\em CORE discussion paper 2003/12 (Accepted by Math. Prog.)}, 2003.

\bibitem[Tib96]{Tibs96}
R.~Tibshirani.
\newblock Regression shrinkage and selection via the lasso.
\newblock {\em Journal of the {R}oyal statistical society, series {B}},
  58(267-288), 1996.

\end{thebibliography}

\end{document}